\documentclass[11pt,epsf]{cernrep}
\usepackage{graphicx}
\usepackage{here}
\input latex.macros
\input top.macros
\def\DZero{D$\mbox{0\kern-.38em /}$}  
\def\invpb{${\rm pb^{-1}}$}

\def\BR{{\cal B}}
\def \mrightdownarrowlong
 {\kern.4em {\vbox{\raise1.25ex \hbox{$|$}
                   \raise2.50ex \hbox{$|$}} \kern-0.27em{
                                                \longrightarrow}}}
\begin{document}
\title{
\begin{flushright}
{\small
CDF/PUB/STATISTICS/PUBLIC/6031 \\
30 June 2002
}
\end{flushright}
Signal Significance in Particle Physics}
\author{Pekka~K.~Sinervo}
\institute{University of Toronto, Toronto, Canada}
\date{21 June 2002}
\maketitle
\begin{abstract}
The concept of the ``statistical significance'' of an observation, and
how it is used in particle physics experiments is reviewed.  
More properly known as a ``p-value,''  the statistical foundations for
this concept are reviewed from a frequentistic perspective.  The
discovery of the top quark at the Fermilab Tevatron Collider and a more
recent analysis of data recorded at Fermilab
are used to illustrate practical applications of these concepts.
\footnote{To be published in the proceedings of the conference on
``Advanced Statistical Techniques in Particle Physics,'' Durham,
England, 18-22 March 2002.}
\end{abstract}

\section{What Particle Physicists Mean by Significance}
When one of your colleagues approaches you and declares that she
has made a ``significant'' observation, intuitively that means that
she has observed some phenomena whose interpretation allows her to
eliminate or falsify one or more hypotheses, and usually support one
or a small number of  alternative hypotheses.  We furthermore expect
that the observation has sufficient statistical power that we expect
that additional observations are unlikely to change these conclusions. 
Scientists have therefore attempted to identify a consistent
statistical framework in which we can quantify this concept of
``significance.''

In particle physics, this concept of statistical significance has not
been employed consistently in the most important discoveries made
over the last quarter century.  Examples of the major discoveries
made over an approximately 10 year period between the late 1970's and
the late 1980's illustrate this point.

Let us consider first the discovery of the
$\Upsilon$\ meson (and the $b$~quark) in 1977 by L.~Lederman and
colleagues \cite{ref: Lederman Upsilon}.  This was made through the
observation of $\mu^+\mu^-$\ final states in high-energy
proton-nucleus collisions at Fermilab, where a large resonant signal
was observed on top of a steeply falling background of dimuon
candidates.  The experimenters estimated that they observed a signal
of approximately 770 events on top of a non-resonant background of
350 candidates.  They characterized the signal as ``significant'' but
made no attempt to quantify or explain exactly what they meant.

The discovery of the $W^-$\ boson at CERN in 1983 by the UA1
collaboration \cite{ref: UA1 W Boson}\ was made by observing 6 events
produced in proton-antiproton collisions where a high energy electron
or antielectron was observed in coincidence with a signature for a
recoiling energetic neutrino.  The collaboration estimated the
background to these 6 events as being ``negligible'' and claimed
discovery of the expected charged weak intermediate vector boson.  This
observation was subsequently confirmed by the UA2 collaboration.

The discovery of $B$\ mesons in 1983 by the CLEO collaboration
\cite{ref: CLEO B Mesons}\ was performed by carefully reconstructing a
variety of different decay modes and illustrating an invariant mass
peak at 5.4~\GeVcc.  The collaboration observed a total
event rate of 17 events on a background of between 4 and 7 events.
They claimed definitive observation of a new particle, but made no
statement that quantified the statistical power of the observation.

As a final example, I note the discovery of $B^0$\ meson flavour mixing
in 1987 by the ARGUS collaboration \cite{ref: ARGUS B Mixing}.  The
experimenters observed $24.8\pm7.6\pm3.8$\ unexpected same-sign
dilepton events versus a total of $25.2\pm5.0\pm3.8$\ opposite-sign
dilepton candidates.  They characterized this as a ``3 $\sigma$''
observation, namely, that the probability that the observed number of
same-sign dilepton events could have been as great or greater than
the observed value was equivalent to the probability of a Gaussian
statistic being observed at least 3 standard deviations from
its expected mean (a probability of $1.35\times 10^{-3}$).

This brief review illustrates that quantifying the statistical 
significance or power  in seminal particle physics measurements is
not uniformly done.  It also illustrates that in at least the one case
in which it was done, the significance was defined as the 
probability\footnote{Unless otherwise noted, ``probability'' in this
article refers to the frequentist definition of this concept.}\ of the
``null hypothesis'' having been responsible for the observation.  

In this paper, I will first review briefly the formal concept of
``statistical significance.''  I will then discuss several examples that
illustrate the use of this concept in particle physics.  I do not have
the opportunity to review all of the techniques that have been in recent
use, but refer the reader to other articles in these proceedings (for
example, the review of the $CL_S$\ method by A.~Read).  

\section{Formal Definitions of Significance}
\subsection{The Frequentists Perspective}
The concept of statistical signficance is formally introduced in the
context of hypothesis testing \cite{ref: Eadie et al}.  Suppose that
we have two hypotheses, $H_0$\ and $H_1$, and a measurement whose
value is a test statistic
$X$\ that, as a random variable, provides some discrimination between
these two hypotheses.  Let $f_0(X)$\ and $f_1(X)$\ represent the
probability distribution functions for $X$\ associated with the two
hypotheses.

Prior to making a measurement of $X$, we would identify a ``critical
region,'' $w$, such that we would select hypothesis $H_1$\ if $X\in
w$\ and $H_0$\ otherwise.  We now have four possible outcomes when we
make a measurement of $X$.  If $X\in w$\ and the hypothesis $H_1$\ is
true, then we have selected the correct hypothesis.  If $X\in w$\ and
$H_0$\ is true, then we have incorrectly concluded that $H_1$\ is
true.  This is known as a mistake of the first kind, and the
probability for this decision is 
\begin{eqnarray}
\int_{X\in w} f_0(X)\, dX &=& \alpha.
\label{eq: significance}
\end{eqnarray}
The probability $\alpha$\ is known as the ``significance'' of the
test.  

\begin{figure}
\begin{center}
\includegraphics[width=\hsize]{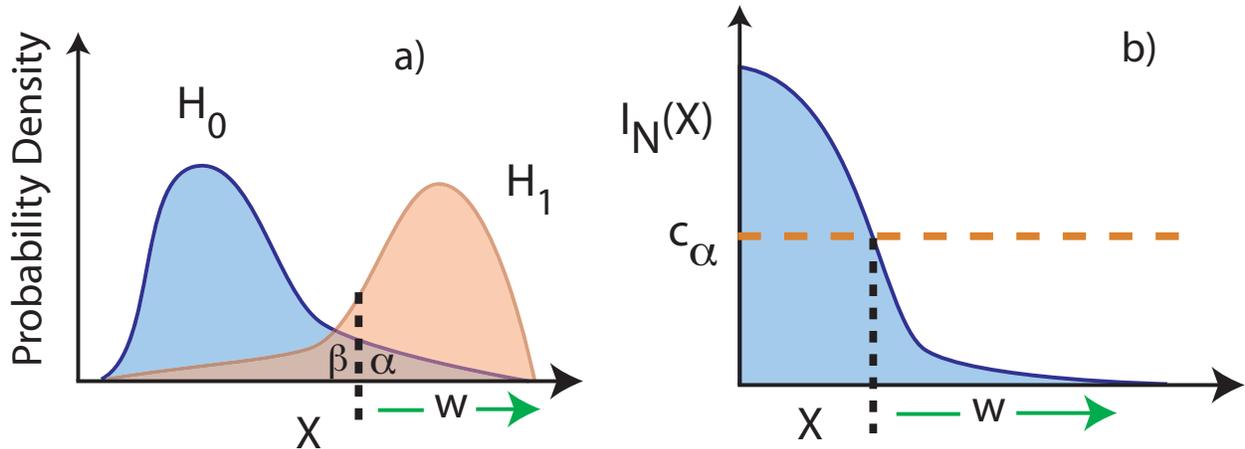}
\end{center}
\caption{A schematic of the hypothesis test described in the text is
shown in a).  The plot shows the probability densities for $X$\ under the
two hypotheses $H_0$\ and $H_1$\ and one possible choice for the region
$w$.  The use of the Neyman-Pearson Theorem is illustrated in b), where
the ratio $I_N$\ is plotted as a function of $X$.}
\label{fig: hypothesis test}
\end{figure}

We have two other possibilities.  The first is if we measure
$X\not\in w$\ when $H_0$\ is true.  In this case, we would have made
the correct inference.  Finally, we have the case where $X\not\in w$\
and $H_1$\ is true.  This is known as a mistake of the second kind,
and the probability for that decision is 
\begin{eqnarray}
\int_{X\not\in w} f_1(X)\, dX &=& \beta.
\end{eqnarray}
The probability $1-\beta$\ is known as the ``power'' of the test. 
The situation is illustrated in Fig.~\ref{fig: hypothesis test}a).
The significance $\alpha$\ is therefore a measure of the ability of a
test to avoid mistakes of the first kind, whereas the power $1-\beta$\
measures the ability of a test to avoid mistakes of the second kind. 
In defining an ``optimimum'' test,  one would like to choose
$X$\ and the region
$w$\ such that $\alpha$\ and $\beta$\ are as small as possible.

\subsection{Significance in Particle Physicists-- The P-Value}
The statistical definition of significance is made in the context of
choosing between two hypotheses.  However, the use of significance in
particle physics discoveries is in a different context. 
The typical case is that an experiment makes a measurement of the
test statistic $X$, say $X_0$.  Furthermore, using the same notation
as before, we assume $X$\ has a probability density $f_0(X)$\ if the
hypothesis $H_0$\ is true.  We further assume that we can categorize
observations of
$X$\ into those that are more and less consistent with $H_0$\
(for the sake of discussion, I will assume that values of $X$\ greater
than
$X_0$\ are less likely given $H_0$).   A measure of the
inconsistency of the observed value
$X_0$\ with the hypothesis
$H_0$\ is then the probability
\begin{eqnarray}
\int_{X>X_0} f_0(X)\, dX.
\end{eqnarray}
This is formally identical to the definition of the
significance in Eq.~\ref{eq: significance}\ if we now define the
critical region to be $w = \{X | X>X_0\}$, i.e. the probability of
observing a value of $X$\ equal to or greater than our observation.
This probability is formally known as the ``p-value'' of the
observation, a convention that the Particle Data Group now has
adopted \cite{ref: PDG 2002}.  
The advantage of using the formal term
for this quantity is that it avoids confusion with the concept of
significance defined in hypothesis tests, where the region $w$\ is
defined {\it a priori}, i.e. before the measurement is made.

The p-value for a given measurement and a specific hypothesis has a
number of features.  First, it only depends on the measurement and the
probability density for the hypothesis.  It is not a
hypothesis test.  It only provides a measure of the consistency of
the hypothesis and the measurement.  In that sense, it is most often
quoted when one has made a measurement that appears to be
inconsistent with a single hypothesis.  A very small p-value is then
used to support the inference that the specific hypothesis should be
rejected.  

Referring then to the example of the discovery of $B^0$\ mixing given in
Section 1, we can now say that the p-value for the observation for the
non-mixing hypothesis was $1.35\times 10^{-3}$.  From a frequentist
perspective, if one rejected the non-mixing hypothesis at this p-value
and it was always true, then one would expect to be wrong (\ie, reject
the correct hypothesis) on average 1 out of every 740 times.

\subsection{A Few More Comments on Hypothesis Testing}
Although the literature uses the p-value of an observation as a
measure of its statistical significance, the concept of
hypothesis testing is an important one in particle physics.  One sees it
most often used in the context where one is designing or proposing an
experiment and wishes to characterize the experiment's ability to
distinguish existing and known physics phenomena (such as that predicted
by the Standard Model) from possible new 
physics \cite{ref: significance uses}.

In those cases, a crucial aspect of the experiment design is the
selection of the optimal statistic $X$\ and the optimal critical
region.  For a specific measurement, such as the observation of a
process above some expected background rate, the choice of $X$\ will
depend on the measurement and the ingenuity of the
experimenter.  One would like to identify test statistics
that have quite different probability density functions for the
hypotheses you wish to distinguish in order to be able to define a
critical region with the smallest possible
$\alpha$\ and
$\beta$.  At the same time, the decision should be informed by the
effect of any systematic uncertainties that may degrade the
separation between two hypotheses.  To that extent, one often
attempts to identify statistics that are not expected to be affected
by systematic uncertainties.  

\subsection{Neyman-Pearson Theorem}
Hypothesis testing has an extensive literature, but relatively few
general results have been identified that can guide our judgement.  One
result, known as the Neyman-Pearson Theorem, is surprisingly useful, and
it is worth reviewing here for the insight it provides.

Suppose we have two hypotheses, $H_0$\ and $H_1$, and we have defined
a test statistic $X$.  Then for a given significance $\alpha$, we can
define the region $w$\ which gives us optimal $\beta$\ (i.e., the
smallest value of $\beta$\ and therefore the greatest statistical
power) by choosing $w$\ as follows. 
We first form the ratio of probability density functions for the two
hypotheses
\begin{eqnarray}
I_N(X) &\equiv& {{f_0(X)}\over{f_1(X)}}.
\end{eqnarray}
The Neyman-Pearson Theorem then concludes that the optimal region
$w$\ is the one over which $I_N(X)$\ is maximal, namely that we
find the value $c_\alpha$\ such that when 
\begin{eqnarray}
w = \{X | I_N(X)< c_\alpha\},
\end{eqnarray}
the probability of observing $X\in w$\ is
\begin{eqnarray}
\int_{X\in w} f_0(X) dX &=& \alpha.
\end{eqnarray}
This construction is illustrated in Fig.~\ref{fig: hypothesis test}b).

The Neyman-Pearson test has one signficant limitation--it is only 
valid for what are known as ``simple hypotheses,'' or hypotheses
where there are no unknown parameters that would be estimated from
the data.  In addition, since it is only applicable when comparing two
hypotheses, it cannot be employed in cases where you have multiple
alternative hypotheses to consider. However, despite these
limitations, this theorem gives us considerable insight into the
definition of the critical region.  For example, we can relate the ratio
of probabilities to the ratio of likelihoods of the two hypotheses:
\begin{eqnarray}
I_N(X) &\equiv& {{f_0(X)}\over{f_1(X)}} \sim {{L_0(X)}\over{L_1(X)}},
\end{eqnarray}
where $L_i(X)$\ are the likelihood functions defined for the two
hypotheses $i=0$\ and $i=1$.
This suggests that the likelihood ratio is one source of guidance for
defining critical regions that have significant (though perhaps not
optimal) power.

\section{The Bayesian Perspective}
Our considerations up to this point have been from a frequentist
perspective, using the standard definition of a frequentist
probability.  In calculating a p-value for a measurement, one has to
assume a hypothesis and then determine the probability (or
probability density) for all possible outcomes of the measurement.  

A Bayesian statistician does not
consider data other than the single measurement.  However, for each
hypothesis, the Bayesian could define a credibility interval that
reflects his or her degree-of-belief in each hypothesis, and the ratio
of these credibility intervals--what is called the ``Bayes discriminant
factor''-- becomes a measure of the relative confidence one has in the
two hypothesis.  Formally, this ratio is
\begin{eqnarray}
{{P(H_0|X)}\over{P(H_1|X)}} =
{{L_0(X)}\over{L_1(X)}}\dot{{\pi_0(X)}\over{\pi_1(X)}},
\end{eqnarray}
where $\pi_0(X)$\ and $\pi_1(X)$\ are the prior probabilities
associated with the two hypotheses.  

This ratio can be used to reject one of the two hypotheses.  The
Bayesians would argue that there is no benefit in attempting to make
anything but a relative statement about the degree-of-belief of the
two hypotheses.  Thus, there is no direct analogy to the p-value in this
framework.  The advantage of this perspective is that it avoids the need
to understand the probability density of all possible outcomes for a
given hypothesis.  It also has the advantage that any inferences you
draw are less sensitive to an outcome that has a low probability
regardless of the hypothesis.  In such cases, the Bayes discriminant
factor still provides information, whereas the p-value is no longer
very informative and could in fact be misleading.  

The disadvantages with this Bayesian approach are,
however, that one has to assume prior distributions for each hypothesis,
and one is only allowed to make relative confidence statements about two
hypotheses.  For these reasons, one finds very limited use of the Bayes
discriminant factor in particle physics.
%
%
\section{P-Values and Experimental Design}
The definition of significance in terms of a p-value for an observation
immediately makes clear the importance of {\it a priori}\ decisions on
the random variables one will measure and how one will define
those observations that prefer one hypothesis over another. 
A carefully designed experiment will identify these and optimize their
choice before any data is analyzed.

However, many particle physics experiments make unique
measurements using general-purpose apparatus designed to study a large
class of processes.  Thus, it is difficult, and often
impossible, to anticipate what one will observe and how.  In fact, early
studies of the data will often guide the experimenters to focus in
specific features that appear unusual or unexpected.  In this context,
the evaluation of a p-value may prove very difficult.  

A simple example illustrates this problem.  Suppose one measures an
invariant mass spectrum in a specific region, say $[m_1, m_2]$,
and one observes a narrow enhancement in a small mass interval, say
$\Delta m$\ wide, of 
$N_o$\ events above an expected background of $N_b$\ events.  In this
case, it would be natural to assume that the hypothesis we wish to
test is the ``null'' hypothesis where we expect $N_b$\ events in this
mass interval $\Delta m$\ and then determine the probability of
observing at least $N_o$\ events.  Assuming that the background rate is
well known (and so we can ignore its uncertainty), the p-value for this
observation would be given by the Poisson probability for observing
at least $N_o$\ events when the mean rate is
$N_b$, or
\begin{eqnarray}
\alpha = \sum_{n=N_o}^\infty
{{\exp\left(-N_b\right)\,\left(N_b\right)^n}\over{n!}}.
\label{eq: simple Poisson estimate}
\end{eqnarray}
However, this probability does not take into account the fact that we
are considering all possible choices of mass interval $\Delta m$\ in the
region $[m_1, m_2]$.  

A proper estimate of this p-value would then
have to include the likelihood of observing at least $N_o$\ events in
{\it any}\ possible interval $\Delta m$.  This 
increases  the p-value of the observation, and changes the possible
inferences one can make.  
For example, a Monte Carlo calculation where
$\delta m$\ is 1\%\ of the interval, $N_o=8$\ events and $N_b=100$\
(i.e., the average number of events in any $\delta m$\ interval is one)
gives a p-value that is 500 times larger than the result in  Eq.~\ref{eq:
simple Poisson estimate}.

\subsection{Blind Analyses}
The prevalence of the p-value in making inferences rests on
the assumption that it is possible to estimate the frequency of
all observations of the test statistic, and that it is possible to
identify the class of observations that are less consistent with a given
hypothesis (the critical region in the language of hypothesis testing). 
This is inherently difficult in cases where one allows the definition of
test statistic and critical region to depend on the actual experimental
outcome itself.  A tactic to eliminate such bias is the ``blind
analysis,'' where one defines the critical region and the statistic
without knowledge of the relevant data \cite{ref: Blind Experiments}.

The ideal experiment is one
in which the measurement and any calculation of its p-value does not have
to be informed by the data itself.
No choices with regard to
selection of data, modifications in the test statistic or choice of
critical region would then be allowed once data collection has started. 
This approach avoids the possibility of selecting,
consciously or unconsciously, a critical region or test
statistic that tend to favour or disfavour a given hypothesis {\it
based on the data observed.}

A number of celebrated failures of inference in particle physics over the
last half-century illustrate what happens when the experimenter allows
the data to guide his or her choices in making inferences about data
\cite{ref: Blind Experiments}.   In all these cases, the quoted p-value
has been assessed incorrectly because it has failed to take into
account how the frequency of a given observation would be affected by
making choices about the test statistic and critical region based on the
actual distribution of the data itself.

\subsection{Use and Limitations of Blind Experiments}
The simple example of ``bump hunting'' illustrates the fundamental
problem in particle physics where one is searching for evidence of new
phenomena;  it is inherently difficult to identify {\it a priori}\ what
class of observations one would expect to use in such a search.  
Besides the difficulty of defining in advance all
possible means of separating ``signal'' from ``background,'' it is also
difficult to limit access to data when one also has to verify that the
instrumentation is working correctly and that any artefacts created by
effects such as miscalibration and
errors in bookkeepping are identified and mitigated.   The experiment
design also has to allow the experimenter access to the data to
measure the rate of background events in the signal sample.  

Despite these challenges, the elimination of certain biases that are
otherwise difficut to control make a blind analysis an attractive
approach given the benefits of being able to make straightforward
estimates of p-values for the possible outcomes.   This technique is
reviewed in another contribution to these proceedings
\cite{ref: Blind Experiments}.

%
%
\section{P-Values in a Counting Experiment}
\subsection{General Considerations}
A common particle physics experiment involves the search for new
phenomena by observing a unique class of events in particle interactions
that cannot be described by background hypotheses.  One usually can
reduce this problem to that of a ``counting experiment,'' where one
identifies a class of events using well-defined criteria, counts
up the total number of observed events, $N_o$, and estimates the average
rate of events, $N_b$, that come from the various background processes. 
One can then perform a straightforward estimate of the p-value of a given
observation of $N_o$\ events, assuming that the probability density for
the random variable $N_o$\ follows a Poisson distribution, \ie\ the
formula in Eq.~\ref{eq: simple Poisson estimate}.

There are several issues that even this simple problem has to address. 
First, one has to be sure that the criteria used to select the class of
events was not in itself biased by how $N_o$\ varied as the criteria
were modified.  Here is where a blind analysis has its greatest benefit,
since this bias is explicitly guarded against.  Second, one has to take
into account possible uncertainties in the estimate of the background
rate $N_b$.  It is beyond the scope of this article to discuss this
issue, and the interested reader is referred to the growing literature
on this topic \cite{ref: uncertainties in backgrounds} (a typical
frequentist approach is to extend the ensemble of possible measurements
to include those experiments with different values of $N_b$\ consistent
with the knowledge of $N_b$).  Third, the careful experimenter has to
make sure that all information relevant to the search is used in the
measurement.  It is at best inefficient and at worst misleading to ignore
relevant data (for example, a possible channel in which the number of
observed events can provide additional information on the process being
studied).   

As a concrete example of the calculation of a p-value for a typical
counting experiment, I will summarize the techniques used by the CDF and
\DZero\ collaborations in their search for top quark production.

\subsection{The Top Quark Search}
The top quark was discovered by pair-production in proton-antiproton
collisions at an energy of 1.8~TeV \cite{ref: CDF top
discovery,ref: DZero top discovery}.  
The top quark decays predominantly
via the process
$t\to W b$, with the
$W$~boson subsequently decaying either leptonically via $W\to l \nu_l$\
(where ``$l$'' can be either an electron, muon or tau lepton) or
hadronically via $W\to q\bar{q}^\prime$\ (the quark final states are
either
$u\bar{d}$\ or
$c\bar{s}$).  This results in three categories of
possible final states with different topologies, efficiencies and
background rates:
\begin{enumerate}
\item the lepton+jets channel, involving one high energy lepton, a
neutrino and three or more jets from the hadronic decay of the $W$\ and
the $b$~quarks,
\item the dilepton channel, involving two high energy leptons, evidence
for two neutrinos, and two or more jets from the $b$~quarks, and
\item the hadronic channel, involving at least six jets.
\end{enumerate}
In both experiments, one had to use
additional criteria to improve the signal-to-noise ratios in the
final candidate event samples.  For CDF, the most effective way to do
this was to require evidence that at least one of the jets arose
from a
$b$~quark using two different ``b-tagging'' techniques.  Thus, one could
characterize the final states by the number of $b$~tags, with the events
with one or two $b$~tags having increasing purity.  For \DZero, the most
effective way to reduce backgrounds was by imposing more stringent
kinematic criteria (a topological selection) and using 
soft muon $b$-tagging.  

The searches used data samples of increasing sensitivity. 
The first reported data came after the
CDF and
\DZero\ collaborations had recorded 19.6 and 15.0 \invpb, respectively
\cite{ref: CDF top 1994,ref: DZero top 1994}.  At that time, the
experiments had not completed analysis of the hadronic channels, which
were expected to be dominated by background.  
The results of
these analyses are summarized in  Table~\ref{tab: top data summary},
where we list the number of observed events, the estimated background
rates, and the branching ratio times efficiency of observing a \ttbar\
decay in each mode. 

\begin{table}
\begin{center}
\begin{tabular}{lcccc}
Final State & Observation  & Expected Background &  
    $\BR\times$~Efficiency & Expected Signal  \\
            &  (events)    &   (events)   &    &   (events)  \\
\hline
\multicolumn{5}{c}{{\bf CDF}} \\
Lepton + Jets (SVX $b$-tags)  & 6   &  $2.3\pm0.3$    &  0.015  &  $2.4$
\\ 
Lepton + Jets (Soft lepton $b$-tags)   & 7   &  $3.1\pm0.3$  & 0.012  
& 
$1.9$
\\ 
Dileptons       &  2   &  $0.6\pm0.3$    & 0.008   &  $1.3$    \\
\hline
\multicolumn{5}{c}{{\bf \DZero}} \\
Lepton + Jets (Soft lepton $b$-tags)   &  2   & $0.6\pm0.2$   & 0.009   &
$1.0$   \\ 
Lepton + Jets (Topology)   &  4   & $1.8\pm0.9$   & 0.026    &
$2.8$   \\ 
Dileptons       &  1  & $0.8\pm0.1$   & 0.007   &  $0.7$  \\
\hline
\end{tabular}
\end{center}
\caption{The observed number of top quark candidates, the expected
background rate, the overall branching ratio times efficiency for the
channel, and the expected number of signal events assuming a top quark
with a mass of 160~\GeVcc\ for each final state.}
\label{tab: top data summary}
\end{table}

The collaborations evaluated the statistical significance of their
data by using a Monte Carlo calculation to estimate the frequency that
the expected background processes would create a combined signal that
was at least as large as that observed.  The Monte Carlo calculation
created an ensemble of experiments that modelled the possible
observations in all channels assuming the Standard Model background
hypothesis.  For a given channel, the estimated background rate was used
as the mean of a Poisson distribution of observed events.  In order to
account for uncertainties in the background rate, the mean value
used to generate a new member of the ensemble was obtained by sampling a
Gaussian distribution with the mean and width of the estimated
background rate
\cite{ref: uncertainties in backgrounds}.  The results of these p-value
calculations are summarized in Table~\ref{tab: top p-values}.  The
collaborations concluded that the individual observations did not provide
sufficient evidence to exclude the background hypothesis.

\begin{table}
\begin{center}
\begin{tabular}{lc}
Final State & P-Value    \\
\hline
\multicolumn{2}{c}{{\bf CDF}} \\
Lepton + Jets (SVX $b$-tagging)  &    0.032 \\
Lepton + Jets (Soft lepton tagging)  &  0.038 \\
Dileptons       & 0.012  \\
Combined       &  0.0026  \\
\hline
\multicolumn{2}{c}{{\bf \DZero}} \\
Combined        &    0.072   \\
\hline
\end{tabular}
\end{center}
\caption{The p-values determined for the observed event rates assuming
the Standard Model background processes by the CDF and \DZero\
collaborations.  The \DZero\ collaboration only reported a p-value for
the observation of 7 candidate events with an expected background of
$3.2\pm1.1$ events.}
\label{tab: top p-values}
\end{table}

The collaborations proceeded to determine how likely their set of
observations were assuming the background hypothesis by identifying a
statistic that combined the observations in the individual channels.  In
the case of a counting experiment involving several channels, the
maximum likelihood estimate of the rate of the process is simply the sum
of the event rates in each channel.  Thus, the natural statistic to
evaluate the combined significance of the observations was the observed
sum of events in all channels.  However, the CDF collaboration noted
that the most sensitive measure of the cross section was not the total
number of observed events in their sample, but the total number of
observed
$b$-tags (since there was a much larger probability of observing
two $b$-tags in a signal event than in an event from a background
process).  Thus, CDF chose as its statistic the sum of the number of
$b$-tags in the lepton+jet events combined and the number of dilepton
events.  Since the \DZero\ \ data relied less on $b$-tagging, the
collaboration chose to use the total number of observed events.

The calculation of the p-value of the observation assuming the
background hypothesis was performed by a Monte Carlo procedure
that effectively created a set of ``pseudo-experiments.''  In
each pseudo-experiment, the number of $b$-tags and dilepton
events from the different background sources was drawn from a Poisson
distribution that had as its mean value the estimated
background rate for the process.  The uncertainty in the various
background components was taken into account as described above, as was
the correlation in the different background sources.  This correlation
arose from the fact that a number of background sources contributed both
types of $b$-tags, whereas others did not.  In effect, this increased the
frequency of observing a larger number of $b$-tags (since now the
fluctuations in the two components were correlated).

The resulting p-values are summarized in Table~\ref{tab: top p-values}. 
One sees that the single most significant p-value was $2.6\times
10^{-3}$.  If one had not taken into account the correlations between
the background sources, the combined p-value would have been $1.6\times
10^{-3}$, or a factor of almost two smaller.  Alternatively, the
combined p-value determined by just counting events would have been
approximately $10^{-2}$.  This demonstrates the sensitivity of a
p-value calculation to the approximations used to determine it.   Given
all this information, both experiments concluded that the observations
were not sufficiently compelling statistically to exclude the background
hypothesis.  

\subsection{Significance Required for Discovery}
In the search for the top quark, the CDF and \DZero\ collaborations
argued that observations with p-values of order $10^{-3}$\ were not
sufficiently significant to be used to claim discovery of a new
phenomenon.  Although this is clearly a matter of opinion, it is roughly
consistent with the practice in the field, where typically the
``$5\sigma$'' standard is used as rough rule of thumb to define the
sensitivity necessary for discovery.  This corresponds to a p-value
equivalent to between $5.7\times 10^{-7}$\ and $2.8\times 10^{-7}$,
depending on whether you are searching for a deviation from a mean or a
one-sided fluctuation from the mean.  
 
As a concrete example, the two Tevatron collaborations used an identical
analysis procedure when approximately a factor of two more data had
been recorded by both experiments.  The resulting p-values
of the CDF and
\DZero\  observations assuming the background hypothesis were $1\times
10^{-6}$\ and $2\times 10^{-6}$, respectively \cite{ref: CDF top
discovery,ref: DZero top discovery}.  Both experiments concluded that
the background hypothesis could be excluded and claimed observation of
top quark pair production.

%
%
\section{P-Values for Continuous Test Statistics}
High-energy physics measurements often examine statistical variables
that are continuous in nature.  In fact, to identify a sample of events
enriched in the signal process, one often imposes selection requirements
on such continuous variables.  Often, it is important to take into
account the entire distribution of a given variable for a set of events,
and not just whether the events lie in a given range of values.  

The general problem can be posed in the following way.  Suppose we have
a set of event data each characterized by a set of statistics
$\vec{X}_i$, where $i=1$\ to $N$.  In addition, one has a hypothesis to
test that predicts the distribution of $\vec{X}$, say
$f(\vec{X};\vec{\alpha})$, where we assume this function to be
normalized to unity between $X_{min}$\ and $X_{max}$, the
minimum and maximum values of $X$, and $\vec{\alpha}$\ is a set of
parameters that are either known or estimated directly from the data. 
Then the general problem is to define a statistic that gives a measure
of the consistency of the distribution of data with the distribution
given by the hypothesis.

\subsection{Possible Tools}
The most widely used such statistic in the 1-dimensional case is a form
of a ``runs test,'' which compares the predicted cumulative distribution
\begin{eqnarray}
g(X) = \int_{X_{min}}^X f(X^\prime)\, dX^\prime
\end{eqnarray}
with the observed cumulative distribution $h(X)$.  The most common
test is the Kolomogorov-Smirnov (K-S) test \cite{ref: KS test}, which
makes this comparison by first finding the K-S distance
\begin{eqnarray}
\delta = \max \left\{ |g(X)-h(X)|, X\in(X_{min},X_{max}) \right\},
\end{eqnarray}
namely the largest difference between the two cumulative distributions.
This test statistic has a characteristic distribution that can be
calculated analytically to provide one with a p-value,
specifically the probability that one would observe a value
of this test statistic as large as or larger than the observed value.  

The K-S test gives a distribution-free measure of the consistency of a
1-dimensional continuous variable and is often used in the
particle physics literature.  Although there are a number of other tests
that could be used in this case, all with similar properties \cite{ref:
Other cumulative tests}, the K-S test has become a reference standard to
employ.  

\subsection{Extension to Higher Dimensions}
The K-S test (and other runs tests) are in principle limited to
1-dimensional distributions, but there are extensions to the case
of several dimensions, though with a number of restrictions.  The
extension requires one to assume that the probability distribution
predicted by the hypothesis can be factorized, so that
\begin{eqnarray}
f(\vec{X}) = f_1(X_1) f_2(X_2) \cdots f_n(X_n),
\end{eqnarray}
where $n$\ is the number of continuous variables being compared.  This
in effect requires each of the variables to be uncorrelated, a strong
assumption and one that has to be verified in practice.  With this
assumption, however, one can then define a set of independent statistics
$\delta_i$, $i=1$\ to $n$, and the associated p-value for each observed
K-S statistic $p_i$.  Then one can combine these independent p-values
into a single measure of significance.  

\subsection{Example:  CDF ``Superjets''}
A concrete example of this technique is a recent analysis of hadron
collider data performed by the CDF collaboration.  A study was performed
of events that were consistent with the production of one or more
hadronic jets and a
$W$\ boson decaying to a lepton-neutrino pair.  The collaboration
defined a subsample of these events where at least one jet was
identified as a ``superjet'', namely a $b$-quark candidate jet with
both the presence of a secondary vertex in the jet displaced from the
interaction vertex and the presence of a second lepton associated with
the jet consistent with coming from the semileptonic decay of a
$b$~hadron \cite{ref: CDF superjets}.  

The collaboration found 13 such events in the 1992-96 Tevatron Collider
data, where they estimated that they would have expected $4.4\pm0.6$\
events from Standard Model background sources.   This observation has a
p-value of 0.001, treating it as a counting experiment and using the
techniques introduced above.   The authors then proceeded to examine nine
separate kinematic variables that had distributions that were predicted
to be  largely uncorrelated, but that might distinguish between the
Standard Model backgrounds and a variety of exotic sources of events.  A
typical example of such a comparison is given in Figure~\ref{fig:
superjet eta}, where the observed distribution of the lepton
pseudorapidity ($\eta
\equiv -\ln\tan(\theta/2)$, where
$\theta$\ is the angle of the lepton relative to the incoming proton
beam axis) is compared with the predicted $\eta$\
distribution.\footnote{The authors chose background distributions
for these figures obtained using Monte Carlo calculations, but used
background distributions for their p-value calculations obtained by
``bootstrapping,'' using a complementary data sample that had no signal
events and that was argued to provide a good characterization of the
expected Standard Model backgrounds.  The Standard Model Monte Carlo
calculations resulted in similar p-value estimates.}\  The plots on the
right-hand side are the distributions of the K-S distance as determined
from a Monte Carlo calculation.  

The p-values from each of the distributions were determined and range
from 0.001 to 0.15.  The authors comment that ``given the {\it a
posteriori}\ selection of the 9 kinematical variables, the combined
statistical significance cannot be unequivocally quantified.''  However,
we can determine a combined p-value by calculating the product
of the 9 p-values, $p_{tot}$, and determining how likely it would be to
obtain this product value assuming the background hypothesis.  This is
given by
\begin{eqnarray}
P_{tot} = \prod_{m=1}^9 \left[ \sum_{k=0}^{m-1} {{-(\ln
p_{tot})^k}\over{k!}} \right],
\end{eqnarray}
and equals $1.6\times 10^{-6}$\ assuming you set aside the
reservations of the authors.

\begin{figure}
\begin{center}
\includegraphics[width=\hsize]{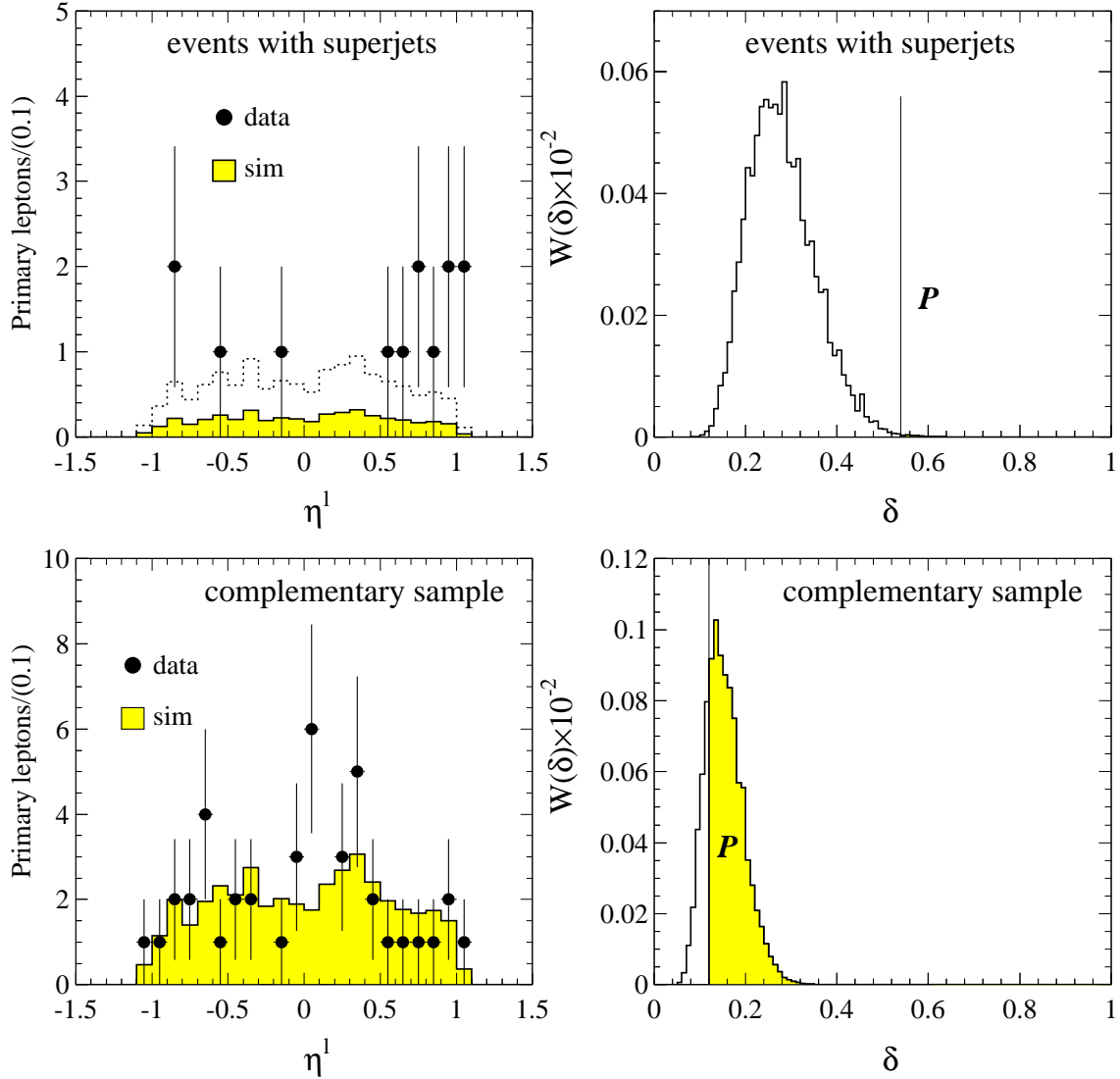}
\end{center}
\caption{The $\eta$\ distribution of the lepton from the $W$\ boson
decay in the CDF ``superjet'' events is shown in the top-left plot
(points) and compared with the Standard Model prediction (shaded
histogram).  The top-right distribution is the expected
distribution of K-S distance of the 13 data events and the SM prediction
in the top-left plot.  The vertical line is the K-S distance for the two
distributions.  Similarly, the bottom-left plot is the lepton $\eta$\
distribution for the complementary sample of data events where a
``superjet'' is not detected, and the bottom-right plot gives the
distribution of the corresponding K-S distance between the data and
predicted distribution.  The K-S test distributions were generated using
a Monte Carlo calculation.}
\label{fig: superjet eta}
\end{figure}

This estimate of the overall p-value raises a number of comments. 
First, are the variables sufficiently uncorrelated that any residual
correlations can be ignored?  Various tests were made of this
assumption by the authors, but no rigorous argument was presented.
Second, uncertainties in the Standard Model predictions have not been
incorporated into the p-value calculation.  These may have some effect
on the overall result, but it is unclear how large this might be.
Third,  the effect on the p-value estimate of the {\it a
posteriori}\ choice of variables is virtually impossible to assess.  A
study of a series of alternate variables were made by the authors, but
no firm conclusion could be drawn.

Of these, perhaps the third is the most vexing.  It is true that the
choice of the 9 variables for this analysis was made after the 13
event data sample had been identified as being unusual.  In that sense,
it is no longer possible to argue that the quoted
p-value is an unbiased
measure of the significance of the observation.  
 
In this case, the best
strategy is to repeat the measurement with an independent data sample to
determine if the same effect is observed.   However, this analysis serves
as a good example of the issues one must face in making such a
multi-variate estimate of significance.

\section{Observations on Current Practice and Summary}
Particle physicists have increasingly relied on numerical estimates of
statistical significance.  The literature is replete with the use of
the p-value, and this appears to have developed into one common measure,
as illustrated by the examples provided above.
Other measures of significance are often quoted, such as the equivalent
number of standard deviations a measurement lies from the value
predicted by a hypothesis.  This is, of course, just a p-value under a
different name.  

More significantly, there are consistent attempts in
the literature to include in p-value estimates more complete information
about a given measurement, such as the sensitivity of the
estimate to systematic uncertainties and information from several
statistics.   The more difficult problem of avoiding unconscious bias in
the selection of statistics is addressed through the use of ``blind
analyses,'' but the effective application of such techniques to
truly serendipitous discoveries is problematic.  Here, the time-honoured
technique of testing specific hypotheses developed through the study of
one data set by creating and analyzing an independent data set with at
least comparable statistical power remains the most effective tool for
separating what we would call the ``statistical fluctations'' from first
evidence for truly new phenomena.

Finally, what is an appropriate criteria for claiming a discovery on the
basis of the p-value of the null hypothesis?  The recent literature 
would suggest a p-value in the range of $10^{-6}$, comparable to
a ``$5\sigma$'' observation, provides convincing evidence.  However, the
credibility of such a claim relies on the care taken to avoid
unconscious bias in the selection of the data and the techniques chosen
to calculate the p-value.

\end{document}